\documentclass[aps,pra,twocolumn,amsmath,amssymb,superscriptaddress,longbibliography]{revtex4-2}
\UseRawInputEncoding
\usepackage[english]{babel}
\usepackage{amssymb}
\usepackage{dcolumn}
\usepackage{bm}
\usepackage{graphicx}
\usepackage{amsmath}
\usepackage{graphicx}       % standard LaTeX graphics tool
\usepackage{mathrsfs}                             % when including figure files
\graphicspath{{pict/}{}}
\usepackage{dcolumn}%%%%%new
\usepackage{bm}
\usepackage[dvipdfm, pdfstartview=FitH, CJKbookmarks=true, bookmarksnumbered=true, bookmarksopen=true, colorlinks=true, pdfborder=001, citecolor=blue, linkcolor=blue, linktocpage=true] {hyperref}
\begin{document}

\title{Unpaired topological triply degenerate point for spin-tensor-momentum-coupled ultracold atoms}

\author{Zhoutao Lei}
\affiliation{Guangdong Provincial Key Laboratory of Quantum Metrology and Sensing $\&$ School of Physics and Astronomy, Sun Yat-Sen University (Zhuhai Campus), Zhuhai 519082, China}
\affiliation{State Key Laboratory of Optoelectronic Materials and Technologies, Sun Yat-Sen University (Guangzhou Campus), Guangzhou 510275, China}

\author{Yuangang Deng}
\email{dengyg3@mail.sysu.edu.cn}
\affiliation{Guangdong Provincial Key Laboratory of Quantum Metrology and Sensing $\&$ School of Physics and Astronomy, Sun Yat-Sen University (Zhuhai Campus), Zhuhai 519082, China}

\author{Chaohong Lee}
\email{lichaoh2@mail.sysu.edu.cn}
\affiliation{Guangdong Provincial Key Laboratory of Quantum Metrology and Sensing $\&$ School of Physics and Astronomy, Sun Yat-Sen University (Zhuhai Campus), Zhuhai 519082, China}
\affiliation{State Key Laboratory of Optoelectronic Materials and Technologies, Sun Yat-Sen University (Guangzhou Campus), Guangzhou 510275, China}

\date{\today}

\begin{abstract}
The realization of triply degenerate points (TDPs) with exotic fermionic excitations has opened a new perspective for the understanding of our nature.
Here we explore the coexistence of single unpaired TDP and multiple twofold Weyl points (WPs) and propose an experimental scheme with ultracold pseudospin-1 atomic gases trapped in optical lattices. We show that the predicted single unpaired TDP emerged by the interplay of quadratic spin-vector- and spin-tensor-momentum-coupling could possess a topological nontrivial middle band. This exotic TDP with mirror symmetry breaking is essential different from the recently observed TDPs that must appear in pairs due to the Nielsen-Ninomiya theorem and host the topological trivial middle band. Strikingly, the topologically protected Fermi-arc states directly connect the unpaired TDP with additional WPs, in contrast to the conventional Fermi-arc states that connect the same degeneracies of band degenerate points with opposite chirality. Furthermore, the different types of TDPs with unique linking structure of Fermi arcs can be readily distinguished by measuring spin texture along high-symmetry lines of the system. Our scheme provides a platform for emerging new fermions with exotic physical phenomena and versatile device applications.
\end{abstract}

\maketitle
\section{introduction\label{Sec1}}
Topological semimetal, a kind of gapless topological phase of matter, is characterized by bulk band degenerate points (BDPs) and supports topologically protected Fermi-arc surface states~\cite{RevModPhys.90.015001,RevModPhys.93.025002}. The Weyl (Dirac) semimetal~\cite{PhysRevB.83.205101,Soluyanov2015, PhysRevLett.108.140405} acts as a paradigm that hosts pairs of twofold (fourfold) linear BDPs. The excitations near these two types of BDPs provide analogs for relativistic fermions with half-integer spins in quantum field theory. Unlike Weyl points (WPs) carrying singly quantized topological charge, Dirac points can be regarded as the overlap of two WPs with opposite chirality. Besides the experimentally realized two- and fourfold BDPs for half-integer spin particles~\cite{Xu613,Lu622,PhysRevX.5.031013, Liu864, PhysRevLett.113.027603, Liu2014, Xu294,Noh2017,PhysRevX.9.021053}, the new types of BDPs with different degeneracies have been studied ranging from condensed matter physics to ultracold quantum gases~\cite{Sun2012NP,Schroter2019,doi:10.1063/1.5124314,doi:10.1126/science.aaz3480,Bradlynaaf5037,PhysRevX.6.031003,PhysRevB.97.121402,PhysRevA.96.033634,PRLCZ2018STMC,PRAZHU2018STMC,PhysRevB.100.235201,Jiang2021}.
These novel fermionic excitations that underlie exotic physical properties are only constrained by the space-group symmetries of the crystal but are not bound by the Poincar\'{e} symmetry imposed in high-energy physics~\cite{PhysRevB.88.104412,PhysRevLett.113.187202, Potter2014,PhysRevX.5.031023,Zhang2016}.

Remarkably, the triply degenerate point (TDP), which represents a new fermion for integer pseudospins without high-energy counterparts~\cite{Bradlynaaf5037,PhysRevX.6.031003,PhysRevA.96.033634,PhysRevB.97.121402,PRLCZ2018STMC,PRAZHU2018STMC,PhysRevB.100.235201,Jiang2021}, has led to tremendous advances in experiments~\cite{Lv2017,Ma2018,PhysRevLett.120.130503}. In general, the typical Hamiltonian of TDPs carrying a double topological charge is made up of the linear order of spin-vector-momentum coupling (SVMC), e.g., $H=\mathbf{q}\cdot\mathbf{F}$, where $\mathbf{{q}}$ is the momentum deviation from the TDPs and $\mathbf{F}$ is the spin-1 matrix. Meanwhile, a different type of spin-tensor-momentum coupling (STMC) can be constructed in high-spin ultracold atoms, which could offer a flexible and controllable platform for exploring exotic topological quantum matters~\cite{PRA88033629,PRLCZ2017STMC,PhysRevA.97.053609,PRACZ2018STMC,PhysRevA.102.013301,PhysRevA.102.013309,Topologicalcoldatoms,RevModPhys.91.015005}.
Recently advances demonstrate that STMC~\cite{PRLCZ2018STMC,TDPZhang} and higher-order dispersion~\cite{PRAZHU2018STMC,PhysRevB.100.235201} spin-1 systems could generate TDPs with different topological charges.

In their pioneer explorations, the TDPs appear in pairs which carry opposite topological charges due to the ``no-g'' theorem~\cite{NIELSEN198120,NIELSEN198122} and host a topological trivial middle band as well. These properties lead to TDPs for integer-spin systems possessing an even number of topological nontrivial bands, which is similar to WPs for half-integer spin. Based on currently available experimental techniques of spin-orbit coupling (SOC)~\cite{NATYJ2011SOCAT,PRLZW2012SOCAT,Ji2014,SCPJW2016SOC,NSHL2016} and Raman-assisted tunneling in optical lattices~\cite{PhysRevLett.107.255301,PhysRevLett.111.185301,PhysRevLett.111.185302,NaJG2014,NPKCJ2015,NPAM2014,science.1259052}, it is of great interest to explore \emph{whether single unpaired TDP with a topological nontrivial middle band can be realized.} An affirmative answer will open up unprecedented opportunities for exploring fundamental particles and provide a broad physics community of applications~\cite{PhysRevB.84.075129,PhysRevB.93.235127,PhysRevLett.118.106402}.

In this work we investigate the coexistence of single unpaired TDP and multiple twofold WPs in a spin-1 system and propose an experimental scheme to realize these exotic TDPs with ultracold atoms trapped in a cubic optical lattice. There appear different types of TDPs induced by quadratic SVMC and STMC.
In particular, a single unpaired TDP hosting a nontrivial middle band is realized with mirror symmetry breaking. We should emphasize that this unconventional TDP has never been observed before and distinguishes the recently observed TDPs appearing in pairs or hosting a topological trivial middle band~\cite{PhysRevA.96.033634,Yang2019,Chang2018,Sanchez2019,doi:10.1126/sciadv.aaw9485,Lv2017,PhysRevLett.120.130503,Ma2018,PRLCZ2018STMC,PRAZHU2018STMC,PhysRevB.97.121402,PhysRevB.100.235201,PhysRevLett.120.130503}.
Moreover, several twofold WPs with single or double topological charge appear, which demonstrates the coexistence of TDP and WPs on certain surfaces.
The coexistence of different types of fermions provides an unprecedented opportunity to significantly enhance our understanding of the properties of basic particles and explore exotic physical phenomena~\cite{PhysRevB.94.165201,PhysRevLett.119.206402,PhysRevLett.120.016401,PhysRevLett.122.076402,PhysRevLett.119.206401,Chang2018,Sanchez2019,Rao2019,PhysRevLett.121.106404,doi:10.1126/sciadv.aaw9485,PhysRevB.98.161403,PhysRevB.99.241104,PhysRevLett.124.105303}.
Remarkably, the nontrivial, topologically protected Fermi arcs connecting different types of fermions (e.g., TDP and WPs) are emerged. In particular, the proposed scheme has the advantage that the generated exotic TDPs can be unambiguously distinguished via spin texture measurements in experiments~\cite{SCPJW2016SOC,NSBS2019,Wang271,TDPZhang}.

This paper is organized as follows. In Sec.~\ref{Sec2} we introduce the model and Hamiltonian for construction of Raman-assisted STMC. Section~\ref{Sec3} is devoted to study the topology of the single unpaired TDPs. In Sec.~\ref{Sec4} we discuss the topologically protected Fermi-arc surface states. In Sec.~\ref{Sec5} we present an accessible measurement method to distinguish the different types of TDPs. Finally, a brief summary is given in Sec.~\ref{Sec6}.

\section{Model and Hamiltonain\label{Sec2}}
\begin{figure}[!htp]
\includegraphics[width=0.85\columnwidth]{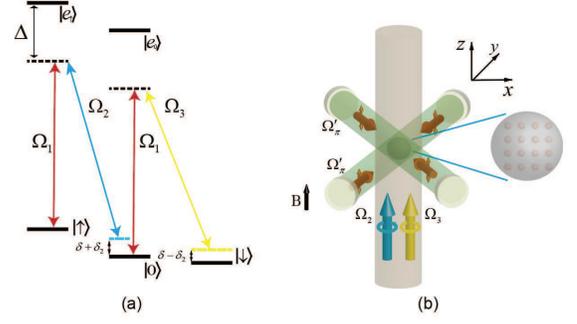}
\caption{\label{model}(a) Involved level structure. (b) Laser configurations for generating quadratic SVMC and STMC.}
\end{figure}

We consider an ultracold gas of $N$ fermionic atoms subjected to a bias magnetic field $\mathbf{B}$ along the quantization $z$ axis. The involved atomic level structure is shown in Fig.~\ref{model}(a), where five relevant magnetic Zeeman sublevels with three ground states and two electronic excited states are involved.
Take $^{40}$K as an example~\cite{PhysRevA.102.013309}; the atomic transition from the ground state $|\sigma\rangle (\sigma=\{\uparrow,0\})$ to the excited state $|e_{\sigma}\rangle$ is driven by a pair of $\pi$-polarized standing-wave lasers with Rabi frequencies $\Omega'_{\pi}\sin(k_Lx-k_Ly)$ and $i\Omega'_{\pi}\sin(k_Lx+k_Ly)$, as shown in Fig.~\ref{model}(b). Here $k_L$ is the wave vector of the laser for the optical lattice with $a=\pi/k_L$ being the lattice constant. Then the total Rabi frequency $\Omega_{1}({\bf r})=\Omega_{\pi}[\sin(k_Lx)\cos(k_Ly)+i\cos(k_Lx)\sin(k_Ly)]$ is generated with $\Omega_{\pi}=\sqrt{2}e^{i\pi/4}\Omega'_{\pi}$~\cite{PhysRevA.95.023611}. To generate SOC, the atomic transition $|0\rangle\leftrightarrow |e_{\uparrow}\rangle$ and $|e_0\rangle\leftrightarrow |\downarrow\rangle$ are driven by two $\sigma$-polarized plane-wave lasers propagating along $z$ direction with the same Rabi frequency $\Omega_2({\bf r})= \Omega_3({\bf r})$.

For large atom-light detuning $\Delta$, two excited states can be adiabatically eliminated to yield a pseudospin-1 system. In the tight-binding regime with a sufficiently strong lattice potential, the lattice Hamiltonian which dominates by the lowest $s$ orbit and nearest-neighbor hopping is given by (see Appendix \ref{appA} for more details)
\begin{eqnarray}\label{single}
{H_0}=\!-\!\!&&\sum_{\mathbf{j},\alpha=x,y}[it_0\hat{\psi}^{\dag}_{\mathbf{j}}\hat{F}_{\alpha}\hat{\psi}_{\mathbf{j} +\mathbf{1}_\alpha} \!+t\hat{\psi}^{\dag}_{\mathbf{j}}(2\hat{F}^2_{z}-\hat{I})\hat{\psi}_{\mathbf{j}+\mathbf{1}_{\alpha}} \!+ {\rm H.c.}]
\nonumber \\
-&&t\sum_{\mathbf{j}}(\hat{\psi}^{\dag}_{\mathbf{j}}\mathcal{R}^{\sigma\sigma}_{z}\hat{\psi}_{\mathbf{j}+\mathbf{1}_z}+ {\rm H.c.})
\nonumber \\
+&&\sum_{\mathbf{j}}[(\delta+\delta_2)\hat{n}_{\mathbf{j},\uparrow}-(\delta-\delta_2)\hat{n}_{\mathbf{j},\downarrow}],
\end{eqnarray}
where $\hat{\psi}_{\mathbf{j}}=[\hat{a}_{\mathbf{j},\uparrow},\hat{a}_{\mathbf{j},0},\hat{a}_{\mathbf{j},\downarrow}]^T$ and $\hat{n}_{\mathbf{j},\sigma}=\hat{a}^{\dag}_{\mathbf{j},\sigma}\hat{a}_{\mathbf{j},\sigma}$, with $\hat{a}_{\mathbf{j},\sigma}$ being the annihilate operator for spin-$\sigma$ component in the $\mathbf{j}$ site, $t_{0}$ is Raman-assisted spin-flip hopping, $t$ is spin-independent hopping, $\mathcal{R}_{z}=\exp(-i\phi\hat{F}_z)$ is induced by the Peierls substitution with Peierls phase $\phi=\sqrt{2}k_La=\sqrt{2}\pi$, and $\delta$ ($\delta_2$) is the effective tunable linear (quadratic) Zeeman shift. Here the unit vectors along three directions are defined by: $\mathbf{1}_x=(1,0,0)$, $\mathbf{1}_y=(0,1,0)$, and $\mathbf{1}_z=(0,0,1)$.

Under the periodic boundary condition with translational invariance, the Hamiltonian (\ref{single}) in momentum space takes the form
\begin{eqnarray}\label{Mo_Hamil}
{H}_0(\mathbf{k})=\sum_{\alpha=x,y,z}d_{\alpha}(\mathbf{k})\hat{F}_{\alpha}+d'_z(\mathbf{k})\hat{F}^2_{z}+\epsilon(\mathbf{k})\hat{I},
\end{eqnarray}
where $\mathbf{k}=(k_x, k_y,k_z)$ is in the first Brillouin zone (FBZ), $d_{x}(\mathbf{k})=2t_0\sin(k_xa)$, $d_{y}(\mathbf{k})=2t_0\sin(k_ya)$, $d_{z}(\mathbf{k})=\delta-t_1\sin(k_za)$, $d'_{z}(\mathbf{k})=\delta_2-4t\cos(k_xa)-4t\cos(k_ya)-t_2\cos(k_za)$, $\epsilon(\mathbf{k})=2t\cos(k_xa)+2t\cos(k_ya)-2t\cos(k_za)$, and $t_1=2t\sin(\sqrt{2}\pi)$ and $t_2=2t[\cos(\sqrt{2}\pi)-1]$ are introduced for shorthand notation. Obviously, the first term $d_{\alpha}(\mathbf{k})\hat{F}_{\alpha}$ represents the conventional SVMC, and the second term $d'_{z}\hat{F}^2_{z}$ denotes the interesting STMC for our high-spin system.

Remarkably, the Hamiltonian~(\ref{Mo_Hamil}) possesses a fourfold rotation symmetry around the $z$ axis $\mathcal{C}_{4z}H_0(k_x,k_y,k_z)\mathcal{C}^{-1}_{4z}=H_0(-k_y,k_x,k_z)$ with $\mathcal{C}_{4z}=e^{-i\pi/2\hat{F}_z}$. As a result, the emerged BDPs will appear at high-symmetry lines due to $\mathcal{C}_{4z}$ symmetry. After ignoring the identity term of $\epsilon(\mathbf{k})\hat{I}$, the interplay between the SVMC $d_{z}(\mathbf{k})\hat{F}_{z}$ and the STMC $d'_z(\mathbf{k})\hat{F}^2_{z}$ will break mirror symmetry, which is necessary to guarantee the appearance of a single unpaired topological TDP. In fact, the separate term of SVMC and STMC is satisfying the different mirror symmetry with ${H}_0(k_x,k_y,k_z)={H}_0(k_x,k_y,\pi-k_z)$ and ${H}_0(k_x,k_y,k_z)={H}_0(k_x,k_y,-k_z)$, respectively. For simplicity, we set $t_0=t$, since the topological phase boundary is independent of nonzero $t_0/t$. Thus the free parameters in our model reduce to the tunable linear Zeeman shift $\delta$ and quadratic Zeeman shift $\delta_2$.

\section{Triply Degenerate Point\label{Sec3}}
To investigate the topology of our system, we calculate the energy spectrum by diagonalizing Hamiltonian (\ref{Mo_Hamil}). The system has three energy bands, whose eigenstates and eigenenergies are respectively denoted as $|\psi_n(\mathbf{k})\rangle$ and $E_n$ with $n=\pm1, 0$ for the $n$th band. In contrast to the previously found TDPs which appear in pairs [e.g., located at ${\bf K}_+ = (0,0,k_0)$ and ${\bf K}_- = (0,0,\pi/a -k_0)$] with opposite chirality~\cite{Lv2017,PhysRevA.96.033634,Ma2018,PhysRevLett.120.130503,PRLCZ2018STMC,PRAZHU2018STMC,PhysRevB.97.121402,PhysRevB.100.235201,PhysRevLett.120.130503} under the ``no-go'' theorem~\cite{NIELSEN198120,NIELSEN198122}, we emphasize that the single TDP in our model is unpaired due to the mirror symmetry breaking. It can be shown that the single unpaired TDP is located at momentum ${\bf K} = (0,0,k_0)$ when the linear and quadratic Zeeman fields satisfy $\delta=t_1\sin(k_0a)$ and $\delta_2=8t+t_2\cos(k_0a)$. Then the effective TDP Hamiltonian with expanding Eq.~(\ref{Mo_Hamil}) in the vicinity of ${\bf K}$ takes the form
\begin{eqnarray}\label{Eq.effectivek0}
H(\mathbf{q})&=&\sum_{\alpha=x,y}v_{\alpha}q_\alpha\hat{F}_{\alpha}+(v_{z}q_z+v'_{z}q^2_z)\hat{F}_{z}\nonumber \\
&&+(u_{z}q_z+u'_{z}q^2_z)\hat{F}^2_{z}+v_0(\mathbf{q})\hat{I},
\end{eqnarray}
where $\mathbf{q}$ indicates the wave vector with respect to ${\bf K}$, $v_{x}=v_{y}=2t_0a$, $v_{z}=-t_1\cos(k_0a)a$, $u_{z}=t_2\sin(k_0a)a$, $v'_{z}=({t_1/2t_2})a u_{z}$, $u'_{z}=-({t_2}/{2t_1})av_z$, and $v_0(\mathbf{q})=4t-2t\cos(k_0a)+2t\sin(k_0a)q_za$. Remarkably, the different types of SOC are generated in our model. Besides the linear dispersion along the $q_z$ axis, both the quadratic dispersion of SVMC and STMC terms emerge. Here, the linear and quadratic Zeeman fields can be treated as the control knobs for SOC strengths $v_{z}$, $u_{z}$, $v'_{z}$, and $u'_{z}$.

In general, the quantized topological charge of a TDP can be characterized by the first Chern number,
\begin{eqnarray}\label{Chern}
\mathcal{C}_n=\frac{1}{2\pi}\oint_{\mathbf{S}}\mathbf{\Omega}_n\cdot d\mathbf{S},
\end{eqnarray}
where $\mathbf{S}$ indicates a closed surface enclosing the TDP and $\mathbf{\Omega}_n(\mathbf{k})=\mathbf{\nabla}_{\mathbf{k}}\times
\langle\psi_n(\mathbf{k})|i\mathbf{\nabla}_{\mathbf{k}}|\psi_n(\mathbf{k})\rangle$ is the Berry curvature of the $n$th band.

\begin{figure}[!htp]
\includegraphics[width=0.95\columnwidth]{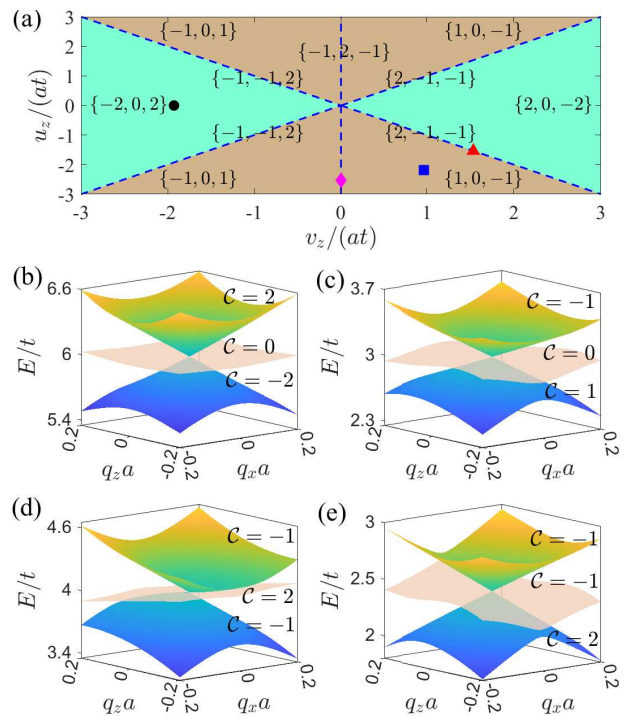}
\caption{\label{TDP}(a) The phase diagram in $v_z$-$u_{z}$ parameter plane. (b)-(e) The energy dispersion near TDP as a function of $q_x$ and $q_z$ with $q_y=0$ for different phases. The black circle and blue square (magenta diamond and red triangle) in (a) mark the parameters for (b) and (c) [(d) and (e)], corresponding to different types of conventional (unconventional) TDPs.}
\end{figure}

In Fig.~\ref{TDP}(a) we show the phase diagram in the $v_z$-$u_{z}$ parameter plane. Each distinct TDP is characterized by a Chern number configuration, $\{\mathcal{C}_{-1},\mathcal{C}_{0},\mathcal{C}_{1}\}$. As can be seen, the Chern number for the $n$th band is $\mathcal{C}_n= -2n v_z/|v_z|$ when the SOC strengths in Eq.~(\ref{Eq.effectivek0}) satisfy $|v_z/u_z|>1$. While for $|u_z/v_z|>1$, the Chern number for the $n$th band is satisfying $\mathcal{C}_n= -n v_z/|v_z|$, which is analogous to generation of topological TDPs induced by STMC~\cite{PRLCZ2018STMC}. We should note that the SOC strength of $v_z$ and $u_{z}$ are highly controlled by $\delta$ and $\delta_2$.

\begin{figure*}[!htp]
\includegraphics[width=1.9\columnwidth]{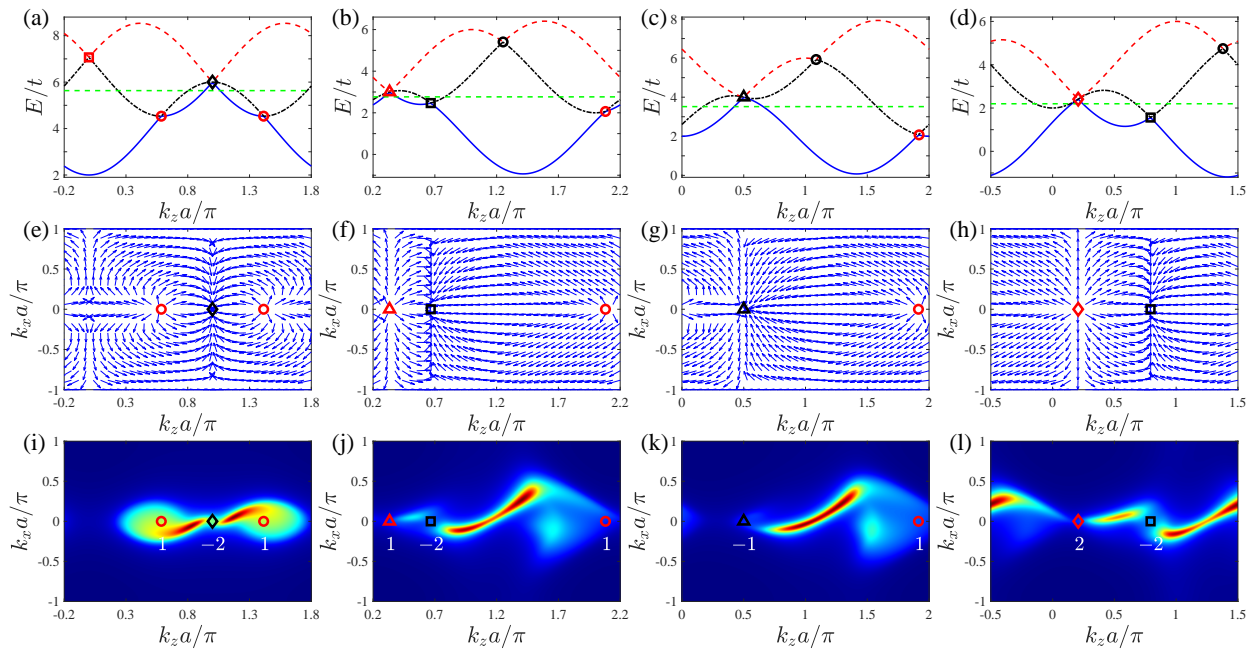}
\caption{\label{energy_cur_fermi}(a)--(d) The energy dispersion along the high-symmetry line $k_x=k_y=0$ for the same parameters as in Figs.~\ref{TDP}(b)--\ref{TDP}(e), respectively. (e--h) The corresponding in-plane Berry curvature of the lowest band $\mathbf{\Omega}_{-1}$ in the $k_y=0$ plane. (i--l) The corresponding surface spectral density on the $(0\overline{1}0)$ surface with Fermi energy shown as a green dashed line in (a--d), respectively. Here the red- (black-) circle (square) and triangle (diamond) denote the single (double) WPs and TDP which hosts a positive (negative) topological charge, respectively. The white numbers represent topological charge of BDPs.}
\end{figure*}

To process further, we use the Chern number of lowest band $\mathcal{C}=\mathcal{C}_{-1}$ to characterize the topological charge of TDP. Therefore the single unpaired TDP carries a double (single) topological charge with $\mathcal{C}=\pm2$ ($\mathcal{C}=\pm1$) in the region $|v_z/u_z|>1$ ($|u_z/v_z|>1$). In Figs.~\ref{TDP}(b) and \ref{TDP}(c), we plot the typical band spectra for the double and single TDP phase, respectively. We show that the middle bands for conventional TDPs are topological trivial with $\mathcal{C}_{0}=0$. These results can be understood as follows. The quadratic terms in Hamiltonian~(\ref{Eq.effectivek0}) do not influence the band closing and opening of bulk gap when $|u_z/v_z|\neq 1$ and $v_z\neq 0$, which indicates the system topology is captured by an effective Hamiltonian with only including linear terms, $\widetilde{H}(\mathbf{q})=\sum_{\alpha=x,y,z}v_{\alpha}q_\alpha\hat{F}_{\alpha}+u_{z}q_z\hat{F}^2_{z}$.
The Hamiltonian $\widetilde{H}(\mathbf{q})$ preserves a linear relation dependence of the momenta with $\widetilde{H}(\mathbf{q})=-\widetilde{H}(-\mathbf{q})$.
Thus the Berry curvature is constrained by the relation $\mathbf{\Omega}_n(\mathbf{q})=\mathbf{\Omega}_{-n}(\mathbf{-q})$,
which yields a topological trivial middle band with $\mathcal{C}_{0}=0$ and $\mathcal{C}_{-1}=-\mathcal{C}_{1}$.
Compared with conventional TDPs, we find three types of unconventional TDPs exhibiting a topological nontrivial middle band (${\cal C}_0\neq 0$) at $v_z=\pm u_{z}'$ or $v_z=0$ [dashed lines in Fig.~\ref{TDP}(a)]. To further characterize these phases, we plot the typical energy spectra $E({\bf q})$ for two of them, as shown in Figs.~\ref{TDP}(d) and \ref{TDP}(e). We emphasize that these unconventional TDPs with ${\cal C}_{0}\neq 0$ are induced by the interplay of quadratic SVMC and STMC. Explicitly, the quadratic SVMC ($v'_{z}q^2_z\hat{F}_{z}$) breaks the relation $\widetilde{H}(\mathbf{q})=-\widetilde{H}(-\mathbf{q})$ at the boundary of topological phase transitions, which yields $\mathcal{C}_{-1}\neq -\mathcal{C}_{1}$. Then the topological nontrivial middle band with nonzero Chern number can be expected since $\sum_n\mathcal{C}_n=0$, which is in contrast to the conventional TDPs proposed in Refs.~\cite{PhysRevA.96.033634,Yang2019,Chang2018,Sanchez2019,doi:10.1126/sciadv.aaw9485,Lv2017,Ma2018,PhysRevLett.120.130503,PRLCZ2018STMC,PRAZHU2018STMC,PhysRevB.97.121402,PhysRevB.100.235201,PhysRevLett.120.130503} that possess topological trivial middle band. In the absence of the STMC ($u_{z}q_z\hat{F}^2_{z}$), we find that the topological nontrivial TDP will reduce to a topological trivial TDP with ${\cal C}_{n}=0$.

\section{Fermi Arcs of TDPs\label{Sec4}}
Besides the unconventional and conventional TDPs, two types of twofold WPs with different topological charges are observed in our system.
Figures~\ref{energy_cur_fermi}(a)--\ref{energy_cur_fermi}(d) show the energy spectrum for different TDPs along the high-symmetry line with $k_x=k_y=0$.
There appear three or two WPs as well as a single unpaired conventional or unconventional TDP in the FBZ, respectively. Furthermore, we plot the Berry curvature of the lowest band $\mathbf{\Omega}_{-1}$ for the four types of TDP phases in the $k_z$-$k_x$ plane, see Figs.~\ref{energy_cur_fermi}(e)--\ref{energy_cur_fermi}(h).
The topological charges of WPs can be well defined by the lower touching band topology. Explicitly, WPs with single ($\mathcal{C}=\pm1$) and double ($\mathcal{C}=\pm2$) topological charges are called single WPs and double WPs, respectively. The emergence of twofold WPs can be characterized by an effective two-band Hamiltonian as discussed in Appendix \ref{appB}. As can be seen, TDP or WPs with positive (negative) Chern number displays as a source (sink) of Berry curvature, which unambiguously demonstrates its nontrivial topological properties. We also check that the total topological charge of all BDPs for the $n$th energy band in the FBZ is zero, that is, $\sum_{BDPs}\mathcal{C}_{n}=0$~\cite{NIELSEN198120,NIELSEN198122}.

To gain more insight into TDPs, we explore the topological protected Fermi arcs, which are the surface states crossing the Fermi energy. Figures~\ref{energy_cur_fermi}(i)--\ref{energy_cur_fermi}(l) display the typical surface spectra density for different types of TDPs on the $(0\overline{1}0)$ surface by using the recursive Green's function method~\cite{Sancho_1985} (see Appendix \ref{appC}). Compared to earlier realized TDPs appearing in pairs~\cite{Lv2017,PhysRevA.96.033634,Ma2018,PhysRevLett.120.130503,PRLCZ2018STMC,PRAZHU2018STMC,PhysRevB.97.121402,PhysRevB.100.235201,PhysRevLett.120.130503}, the emerged single TDP is unpaired due to mirror symmetry breaking. Interestingly, in our system the Fermi-arc states will link the single unpaired TDP and multiple WPs, which is essentially different from the observed coexistence of paired TDPs or WPs where Fermi arcs link the same types of fermions in materials~\cite{Lv2017,https://doi.org/10.1002/pssr.201900421}. It clearly shows that each single unpaired TDP has a unique linking structure of Fermi arcs.

We find that the number of Fermi arcs linking to a single unpaired TDP is determined by its topological charge $\mathcal{C}$. Moreover, the Fermi arcs directly link single unpaired TDPs with two (one) WPs for conventional (unconventional) TDP phases, as shown in Figs.~\ref{energy_cur_fermi}(i) and \ref{energy_cur_fermi}(j) [Figs.~\ref{energy_cur_fermi}(k) and \ref{energy_cur_fermi}(l)]. It is shown that the conventional TDPs with $\mathcal{C}=-2$ and $\mathcal{C}=1$ are both exhibiting two separate Fermi arcs with three nodes. These Fermi-arc patterns [Figs.~\ref{energy_cur_fermi}(i) and \ref{energy_cur_fermi}(j)] linking single unpaired TDP with two WPs are yet to be studied, which is in contrast to the recently study in Refs.~\cite{PhysRevLett.119.206402,Chang2018,PhysRevLett.120.016401,Sanchez2019,PhysRevLett.122.076402,doi:10.1126/sciadv.aaw9485,Yang2019}
that Fermi arcs link the same number of conventional TDPs with fourfold BDPs with opposite chirality. Of particular interest, the unconventional TDP with single (double) topological charge connects with one WP with opposite chirality and hosts one (two) Fermi arcs as well, as displayed in Fig.~\ref{energy_cur_fermi}(k) [Fig.~\ref{energy_cur_fermi}(l)]. In fact, an unconventional TDP can be treated as a combination of one conventional TDP and one WP (see Appendix \ref{appD}), which could provide a new insight into unconventional TDP with a topological nontrivial middle band. Remarkably, the observed exotic Fermi arcs could act as the unique surface fingerprints for different types of TDPs.

\begin{figure}[!htp]
\includegraphics[width=0.95\columnwidth]{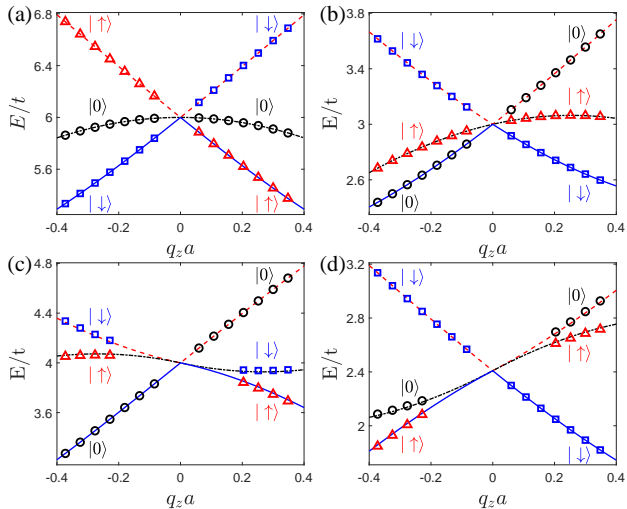}
\caption{\label{spin_energy}The spin textures for different energy bands near a TDP along $k_x=k_y=0$ for spin-$\uparrow$ ({\color{red}$\bigtriangleup$}), spin-$0$ ($\bigcirc$), and spin-$\downarrow$ ({\color{blue}$\Box$}). The parameters are the same as in Figs.~\ref{TDP}(b)--(e).}
\end{figure}

\section{Determination of TDPs\label{Sec5}}
In order to further reveal the topology of different TDPs, we calculate their spin textures. We should emphasize that the eigenstates of helicity branches for Hamiltonian~(\ref{Mo_Hamil}) are always reduced to bare spin states at the high-symmetry line $k_x=k_y=0$  due to $\mathcal{C}_{4z}$ symmetry. These properties provide a unique opportunity to experiment measure the Chern number of single unpaired TDPs through spin-resolved time-of-flight images in ultracold quantum gases~\cite{SCPJW2016SOC,NSBS2019,Wang271}. Figure~\ref{spin_energy} shows the typical spin textures around four types of TDP phases at $k_x=k_y=0$. As can be seen, the bare spin state of the middle band does not cause spin flips at the band crossing for conventional double [Fig~\ref{spin_energy}(a)] and single [Fig~\ref{spin_energy}(b)] TDP, respectively, which is consistent with the results of predicting a topological trivial middle band, as displayed in Figs.~\ref{TDP}(b) and \ref{TDP}(c). Interestingly, the bare spin state of the lowest band will cause spin flips during the band crossing process. The double (single) spin flips corresponding to the spin-angular-momentum transition from $|\downarrow\rangle$ ($|0\rangle$) to $|\uparrow\rangle$ ($|\downarrow\rangle$) is in connection with topological double (single) TDPs with $\mathcal{C}=-2$ ($\mathcal{C}=1$).

Compared with the conventional TDPs, the bare spin state of the middle band for unconventional TDPs is observed in the spin flips process, as shown in Figs.~\ref{spin_energy}(c) and \ref{spin_energy}(d). Remarkably, the Chern number for the topological nontrivial middle band is equivalent to changing values of spin angular momentum after spin flips. Analogously, the single and double TDPs correspond to single ($|0\rangle \rightarrow |\uparrow\rangle$) and double ($|\uparrow\rangle \rightarrow |\downarrow\rangle$) spin flips for the lowest band, respectively. Remarkably, both conventional and unconventional TDPs with nonzero topological charges can be directly extracted by measuring spin-flip processes along the high-symmetry line in experiments~\cite{SCPJW2016SOC,NSBS2019,Wang271,TDPZhang}. The advantage of our proposal is that the characterization of these novel topological TDPs do not need to measure the Chern number of bands~\cite{NPAM2014,science.1259052} and Bloch state tomography over the FBZ for ultracold quantum gasses~\cite{science.aad5812,science.aad4568}. In addition, the generation of Raman-assisted STMC can utilize the self-ordering dynamic SOC in cavity quantum electrodynamics~\cite{PhysRevLett.112.143007,PhysRevLett.121.163601,PhysRevLett.123.160404}, where there is an additional collective-emission-induced cooling mechanism which will facilitate the experimental realization of many-body topological quantum matters. Finally, we should emphasize that the Raman-induced heating can be greatly mitigated by employing the long-lived excited states in alkali-earth atoms~\cite{PhysRevLett.116.035301,PhysRevLett.117.220401,PhysRevX.6.031022}.

\section{Conclusions\label{Sec6}}
Based upon the currently available techniques of Raman-induced SOC, we explore the generation of exotic TDPs beyond the Nielsen-Ninomiya theorem and discuss its experimental realization via ultracold pseudospin-1 atoms. We give the phase diagram via analyzing the topological properties of these exotic TDPs. It has been shown that the interplay of the quadratic SVMC and STMC gives rise to the coexistence of topological single unpaired TDP and multiple twofold WPs. In particular, the unconventional single and double TDPs with a topological nontrivial middle band are observed. Interestingly, there emerge four different patterns of topologically protected Fermi-arc states directly linking TDPs and WPs, which corresponds to the coexistence of TDPs and WPs. As to the experimental detection, an accessible measurement method to distinguish different types of TDPs was proposed, which could facilitate the experimental feasibility for studying topological semimetal in cold atoms. Finally, we point out that our scheme can be extended to study exotic topological Fulde-Ferrell superfluid in high-spin systems~\cite{PhysRevA.95.023611,zhang2013topological,PhysRevLett.113.115302,PhysRevLett.112.136402}.

\begin{acknowledgments}
{This work is supported by the National Key R$\&$D Program of China (Grant No. 2018YFA0307500), the NSFC (Grants No. 11874433, No. 12135018, No. 11874434, and No. 12025509), the Key-Area Research and Development Program of GuangDong Province under Grant No. 2019B030330001, and the Science and Technology Program of Guangzhou (China) under Grant No. 201904020024.}
\end{acknowledgments}
%\acknowledgement
\appendix

\section{Lattice Hamiltonian\label{appA}}
%%%%%%%%%%%%%%%%%%%%%%%%%%%%%%%%%%%%%%%%%%%%%%%%%%%%%%%%%%%%%%%%%%%%%%%%%%%%%%%%%%%%%%%%%%%%%%%%%%%%%%%%%%%%%%%%%%%%%%%%%%%%%%%%%%%%%%%
%\setcounter{figure}{0}
In this section we derive the lattice Hamiltonian in detail with the level diagram and laser configuration given in Fig.~\ref{model} of the main text. %The atomic structure with Hamiltonian about ground states and involved existed states will be ($\hbar=1$)
%\begin{eqnarray}\label{Eq.Ham at}
%\boldsymbol{h}_{in}=&\omega_Z&\hat{b}^{\dag}_{\uparrow}\hat{b}_{\uparrow}+\omega_q \hat{b}^{\dag}_{0}\hat{b}_{0}-\omega_Z\hat{b}^{\dag}_{\downarrow}\hat{b}_{\downarrow} \nonumber \\
%&+&(\omega_a+\omega'_Z)\hat{e}^{\dag}_{\uparrow}\hat{e}_{\uparrow}+(\omega_a+\omega'_q)\hat{e}^{\dag}_{0}\hat{e}_{0},
%\end{eqnarray}
%where $\hat{b}_{\sigma=\uparrow,0,\downarrow}$ and $\hat{e}_{\sigma=\uparrow,0}$ are, respectively, the annihilation operators for ground and excited states. The atomic transition frequency for ground states to excited states is indicated by $\omega_a$. $\omega_Z$ and $\omega_q$ ($\omega'_Z$ and $\omega'_q$) are the linear and quadratic Zeeman shift for the ground states (excited states), respectively.
As shown in Fig.~\ref{model} of main text, the atomic transition from the ground state $|\sigma\rangle (\sigma=\{\uparrow,0\})$ to the excited state $|e_{\sigma}\rangle$
is driven by a pair of $\pi$-polarized standing-wave lasers with Rabi frequencies $\Omega'_{\pi}\sin(k_Lx-k_Ly)$ and $i\Omega'_{\pi}\sin(k_Lx+k_Ly)$, corresponding the atom-light detuning $\Delta=\omega_a-\omega_L$, with $\omega_L$ being the laser detuning. Then the total Rabi frequency $\Omega_{1}({\bf r})=\Omega_{\pi}[\sin(k_Lx)\cos(k_Ly)+i\cos(k_Lx)\sin(k_Ly)]$ is generated with $\Omega_{\pi}=\sqrt{2}e^{i\pi/4}\Omega'_{\pi}$.
To generate spin-orbit coupling (SOC), the atomic transitions $|0\rangle\leftrightarrow |e_{\uparrow}\rangle$ and $|e_0\rangle\leftrightarrow |\downarrow\rangle$
are driven by $\sigma$-polarized plane-wave lasers propagating along the $z$ direction with the same Rabi frequency $\Omega_2({\bf r})= \Omega_3({\bf r})=\Omega_{\sigma}e^{i\sqrt{2}k_Lz}$, where the laser frequency is $\omega_L+\Delta\omega_L$ and $\omega_L-\Delta\omega_L'$, respectively. Then the single-particle Hamiltonian is given by
\begin{eqnarray}\label{Eq.Hamin at}
\boldsymbol{h}_{in}=&\omega_Z&\hat{b}^{\dag}_{\uparrow}\hat{b}_{\uparrow}+\omega_q \hat{b}^{\dag}_{0}\hat{b}_{0}-\omega_Z\hat{b}^{\dag}_{\downarrow}\hat{b}_{\downarrow} \nonumber \\
&+&(\omega_a+\omega'_Z)\hat{e}^{\dag}_{\uparrow}\hat{e}_{\uparrow}+(\omega_a+\omega'_q)\hat{e}^{\dag}_{0}\hat{e}_{0} \nonumber \\
&+&[e^{i\omega_Lt}(\mathbf{\Omega^*_{1}}\hat{b}^{\dag}_{\uparrow}\hat{e}_{\uparrow}+\mathbf{\Omega^*_{1}}\hat{b}^{\dag}_{0}\hat{e}_{0})+{\rm H.c.}] \nonumber \\
&+&[\mathbf{\Omega^*_{2}}e^{i(\omega_L+\Delta\omega_L)t}\hat{b}^{\dag}_{0}\hat{e}_{\uparrow}+{\rm H.c.}] \nonumber \\
&+&[\mathbf{\Omega^*_{3}}e^{i(\omega_L-\Delta\omega_L')t}\hat{b}^{\dag}_{\downarrow}\hat{e}_{0}+{\rm H.c.}],
\end{eqnarray}
where $\hat{b}_{\sigma=\uparrow,0,\downarrow}$ and $\hat{e}_{\sigma=\uparrow,0}$ are, respectively, the annihilation operators for the ground and excited states. The atomic transition frequency for ground states to excited states is indicated by $\omega_a$. $\omega_Z$ and $\omega_q$ ($\omega'_Z$ and $\omega'_q$) are the linear and quadratic Zeeman shift for the ground states (excited states), respectively.

By introducing the rotating frame that is defined by the unitary transformation $\tilde{\mathcal{U}}=e^{-iUt}$ with
\begin{eqnarray}\label{Eq.Unita}
U=&&(\omega_q+\Delta\omega_L)\hat{b}^{\dag}_{\uparrow}\hat{b}_{\uparrow}+\omega_q\hat{b}^{\dag}_{0}\hat{b}_{0}
\nonumber \\
&&+(\omega_q+\Delta\omega_L')\hat{b}^{\dag}_{\downarrow}\hat{b}_{\downarrow} \nonumber \\
&&+(\omega_q+\Delta\omega_L+\omega_L)\hat{e}^{\dag}_{\uparrow}\hat{e}_{\uparrow}+(\omega_q+\omega_L)\hat{e}^{\dag}_{0}\hat{e}_{0},
\end{eqnarray}
then the Hamiltonian (\ref{Eq.Hamin at}) reduces to \begin{eqnarray}\label{Eq.Rotating}
\boldsymbol{h}_{in}&\rightarrow&\tilde{\mathcal{U}}^\dagger\boldsymbol{h}_{in}\tilde{\mathcal{U}}-i\tilde{\mathcal{U}}^\dagger\frac{\partial}{\partial_t}\tilde{\mathcal{U}}
\nonumber \\
&=&\delta_{\uparrow}\hat{b}^{\dag}_{\uparrow}\hat{b}_{\uparrow}-\delta_{\downarrow}\hat{b}^{\dag}_{\downarrow}\hat{b}_{\downarrow}
\nonumber \\
&+&(\Delta+\delta_{\uparrow}-\omega_z+\omega'_z)\hat{e}^{\dag}_{\uparrow}\hat{e}_{\uparrow}+(\Delta-\omega_q+\omega'_q)\hat{e}^{\dag}_{0}\hat{e}_{0} \nonumber \\
&+&[\mathbf{\Omega^*_{1}}\hat{b}^{\dag}_{\uparrow}\hat{e}_{\uparrow}+\mathbf{\Omega^*_{1}}\hat{b}^{\dag}_{0}\hat{e}_{0}+{\rm H.c.}] \nonumber \\
&+&[\mathbf{\Omega^*_{2}}\hat{b}^{\dag}_{0}\hat{e}_{\uparrow}+\mathbf{\Omega^*_{3}}\hat{b}^{\dag}_{\downarrow}\hat{e}_{0}+{\rm H.c.}].
\end{eqnarray}
Here $\delta_{\uparrow}=\omega_z-\omega_q-\Delta\omega_L$ and $\delta_{\downarrow}=\omega_z+\omega_q+\Delta\omega'_L$ are the two-photon detuning. As a result, we can get the Heisenberg equations about different internal states
\begin{eqnarray}\label{Eq.Heisenberg}
i\frac{\partial\hat{b}_{\uparrow}}{\partial t}&=&\delta_{\uparrow}\hat{b}_{\uparrow}+\mathbf{\Omega^*_{1}}\hat{e}_{\uparrow} , \nonumber \\
i\frac{\partial\hat{b}_{0}}{\partial t}&=&\mathbf{\Omega^*_{1}}\hat{e}_{0}+\mathbf{\Omega^*_{2}}\hat{e}_{\uparrow}, \nonumber \\
i\frac{\partial\hat{b}_{\downarrow}}{\partial t}&=&-\delta_{\downarrow}\hat{b}_{\downarrow}+\mathbf{\Omega^*_{3}}\hat{e}_{0}, \nonumber \\
i\frac{\partial\hat{e}_{\uparrow}}{\partial t}&=&(\Delta+\delta_{\uparrow}-\omega_z+\omega'_z-i\gamma)\hat{e}_{\uparrow}+\mathbf{\Omega_{1}}\hat{b}_{\uparrow}+\mathbf{\Omega_{2}}\hat{b}_{0}, \nonumber \\
i\frac{\partial\hat{e}_{0}}{\partial t}&=&(\Delta-\omega_q+\omega'_q-i\gamma)\hat{e}_{0}+\mathbf{\Omega_{1}}\hat{b}_{0}+\mathbf{\Omega_{3}}\hat{b}_{\downarrow},
 \end{eqnarray}
where $\gamma$ is the spontaneous emission rate for the excited states. For the large detuning limit, i.e., $|\Omega_{\sigma,\pi}/\Delta|\ll1$, $|\delta_{\uparrow,\downarrow}/\Delta|\ll1$, and $|\gamma/\Delta|\ll1$, we can adiabatically eliminate the exited states, i.e., $i\dot{\hat{e}}_{\uparrow,0}=0$, which yields
\begin{eqnarray}\label{Eq.exited}
\hat{e}_{\uparrow}\approx&&-\frac{\mathbf{\Omega_{1}}\hat{b}_{\uparrow}+\mathbf{\Omega_{2}}\hat{b}_{0}}{\Delta}, \nonumber \\
\hat{e}_{0}\approx&&-\frac{\mathbf{\Omega_{1}}\hat{b}_{0}+\mathbf{\Omega_{3}}\hat{b}_{\downarrow}}{\Delta}.
\end{eqnarray}
After substituting Eq.~(\ref{Eq.exited}) to Eq.~(\ref{Eq.Heisenberg}), we obtain the effective Heisenberg equation about ground states:
\begin{eqnarray}\label{Eq.ground}
i\frac{\partial\hat{b}_{\uparrow}}{\partial t}=&&\delta_{\uparrow}\hat{b}_{\uparrow}-\frac{1}{\Delta}\mathbf{\Omega^*_{1}}\mathbf{\Omega_{2}}\hat{b}_{0},
 \nonumber \\
i\frac{\partial\hat{b}_{0}}{\partial t}=&&-\frac{1}{\Delta}[\mathbf{\Omega_{1}}\mathbf{\Omega^*_{2}}\hat{b}_{\uparrow}
+|\mathbf{\Omega_{2}}|^2\hat{b}_{0}
+\mathbf{\Omega^*_{1}}\mathbf{\Omega_{3}}\hat{b}_{\downarrow}], \nonumber \\
i\frac{\partial\hat{b}_{\downarrow}}{\partial t}=&&-\delta_{\downarrow}\hat{b}_{\downarrow}-\frac{1}{\Delta}[\mathbf{\Omega_{1}}\mathbf{\Omega^*_{3}}\hat{b}_{0}
+|\mathbf{\Omega_{3}}|^2\hat{b}_{\downarrow}].
\end{eqnarray}
Here we assume the Raman coupling strength of the standing wave is sufficiently weak with satisfying $\Omega_{\pi}/\Omega_{\sigma}\ll 1$. Therefore the $\pi$-polarized laser-induced position-dependent Stark shift can be safely ignored. Then the effective Hamiltonian for ground-states space is given by
\begin{eqnarray}\label{Eq.groundHa}
\boldsymbol{h}_{in}=&&\left(
  \begin{array}{ccc}
    \delta+\delta_2 & -\frac{1}{\Delta}\mathbf{\Omega^*_{1}}\mathbf{\Omega_{2}} & 0\\
    -\frac{1}{\Delta}\mathbf{\Omega_{1}}\mathbf{\Omega^*_{2}} & 0 & -\frac{1}{\Delta}\mathbf{\Omega^*_{1}}\mathbf{\Omega_{3}}\\
    0 & -\frac{1}{\Delta}\mathbf{\Omega_{1}}\mathbf{\Omega^*_{3}} & -\delta+\delta_2
  \end{array}
\right),
\end{eqnarray}
where $\delta=\frac{\delta_{\uparrow}+\delta_{\downarrow}}{2}+\frac{|\Omega_{\sigma}|^2}{2\Delta}$  ($\delta_2=\frac{\delta_{\uparrow}-\delta_{\downarrow}}{2}-\frac{3|\Omega_{\sigma}|^2}{2\Delta}$) is the effective tunable linear (quadratic) Zeeman shift.

Incorporating the center-of-mass motion, the system Hamiltonian reads
\begin{eqnarray}\label{Eq.groundHa}
\boldsymbol{h}_0=\left[\frac{\mathbf{p}^2}{2M}+\mathcal{U}_{ol}(\mathbf{r})\right]\hat{I}+\boldsymbol{h}_{in},
\end{eqnarray}
where $M$ is the atomic mass, and $\hat{I}$ is the identity matrix. After applying the gauge transformation $\{|\uparrow\rangle\rightarrow e^{-i\sqrt{2}k_Lz}|\uparrow\rangle, |\downarrow\rangle\rightarrow e^{i\sqrt{2}k_Lz}|\downarrow\rangle\}$, the effective single-particle atom-light Hamiltonian becomes
\begin{eqnarray}
{\boldsymbol{h}}_0=&&\frac{({\mathbf{p}}-{\mathbf{A}})^{2}}{2M}+M_x\mathbf{(r)}\hat{F}_{x}+M_y\mathbf{(r)}\hat{F}_{y}\nonumber \\
&&+\delta\hat{F}_{z}+\delta_2\hat{F}^2_z+\mathcal{U}_{\rm{ol}}(\mathbf{r})\hat{I},
\end{eqnarray}
where $\hat{F}_{x,y,z}$ is the spin-1 matrix, ${\mathbf{A}}=\sqrt{2}\hbar k_L\hat{F}_z$ is the vector potential introduced by the gauge transformation, $M_x(r)=\Omega\sin(k_Lx)\cos(k_Ly)$, and $M_y(r)=\Omega\cos(k_Lx)\sin(k_Ly)$ with $\Omega=-\sqrt{2}\Omega_{\pi}\Omega_{\sigma}/\Delta$.

Furthermore, we consider the atoms are deeply trapped by a spin-independent red-detuned three-dimensional ($3$D) optical lattice $\mathcal{U}_{\rm ol}(\mathbf{r})=-U_{\rm ol}[\cos^2(k_Lx)+\cos^2(k_Ly)+\cos^2(k_Lz)]$. Here $U_{\rm ol}$ is the depth of lattice, and $k_L$ is the wave vector of the optical trapping laser with the lattice constant $a=\pi/k_L$. When the lattice potential is sufficiently strong, the system enters the tight-binding regime. By only considering the lowest $s$ orbits and nearest-neighbor hopping, the lattice Hamiltonian takes the form as
%\begin{widetext}
%\begin{eqnarray}\label{Eq.Lattice}
%{H_0}=~&&\sum_{\sigma=\uparrow,\downarrow}\sum_{\mathbf{j},\zeta=\pm1}(-1)^{(m+n)}\frac{t_0}{\sqrt{2}}(\zeta\hat{a}^{\dag}_{\mathbf{j},\sigma}\hat{a}_{\mathbf{j}+\zeta\mathbf{1}_x,0}
%-i\zeta\hat{a}^{\dag}_{\mathbf{j},\sigma}\hat{a}_{\mathbf{j}+\zeta\mathbf{1}_y,0}+{\rm H.c.})
%\nonumber \\
%-&&\sum_{\sigma=\uparrow,0,\downarrow}\sum_{\mathbf{j}}t(\hat{a}^{\dag}_{\mathbf{j},\sigma}\hat{a}_{\mathbf{j}+\mathbf{1}_{x},\sigma}+\hat{a}^{\dag}_{\mathbf{j},\sigma}\hat{a}_{\mathbf{j}+\mathbf{1}_{y},\sigma}+\hat{a}^{\dag}_{\mathbf{j},\sigma}\mathcal{R}^{\sigma\sigma}_{z}\hat{a}_{\mathbf{j}+\mathbf{1}_{z},\sigma}+{\rm H.c.})
%\nonumber \\
%+&&\sum_{\mathbf{j}}[(\delta+\delta_2)\hat{n}_{\mathbf{j},\uparrow}-(\delta-\delta_2)\hat{n}_{\mathbf{j},\downarrow}],
%\end{eqnarray}
%\end{widetext}
\begin{widetext}
\begin{eqnarray}\label{Eq.Lattice}
{H_0}=~&&\sum_{\sigma=\uparrow,\downarrow}\sum_{\mathbf{j},\zeta=\pm1}(-1)^{(m+n)}\frac{t_0}{\sqrt{2}}(\zeta\hat{a}^{\dag}_{\mathbf{j},\sigma}\hat{a}_{\mathbf{j}+\zeta\mathbf{1}_x,0}
-i\zeta\hat{a}^{\dag}_{\mathbf{j},\sigma}\hat{a}_{\mathbf{j}+\zeta\mathbf{1}_y,0}+{\rm H.c.}) + \sum_{\mathbf{j}}[(\delta+\delta_2)\hat{n}_{\mathbf{j},\uparrow}-(\delta-\delta_2)\hat{n}_{\mathbf{j},\downarrow}]
\nonumber \\
-&&\sum_{\sigma=\uparrow,0,\downarrow}\sum_{\mathbf{j}}t(\hat{a}^{\dag}_{\mathbf{j},\sigma}\hat{a}_{\mathbf{j}+\mathbf{1}_{x},\sigma}+\hat{a}^{\dag}_{\mathbf{j},\sigma}\hat{a}_{\mathbf{j}+\mathbf{1}_{y},\sigma}+\hat{a}^{\dag}_{\mathbf{j},\sigma}\mathcal{R}^{\sigma\sigma}_{z}\hat{a}_{\mathbf{j}+\mathbf{1}_{z},\sigma}+{\rm H.c.}),
\end{eqnarray}
\end{widetext}
where $\hat{a}_{\mathbf{j},\sigma}$ is the annihilate operator for the $\sigma$ component in the $\mathbf{j}$ site, and $\hat{n}_{\mathbf{j},\sigma}=\hat{a}^{\dag}_{\mathbf{j},\sigma}\hat{a}_{\mathbf{j},\sigma}$. Here, we define the 3D lattice index $\mathbf{j}\equiv{(m,n,l)}$ and unit vectors along three directions:
$\mathbf{1}_x=(1,0,0)$, $\mathbf{1}_y=(0,1,0)$, and $\mathbf{1}_z=(0,0,1)$.
Furthermore, $t=-\int d{\mathbf{r}}w^*_{\mathbf{i}}({\mathbf{r}})[{\mathbf{p^2}}/(2M)+\mathcal{U}_{ol}(\mathbf{r})]w_{\mathbf{i}+\mathbf{1}_\alpha}({\mathbf{r}})$ is the spin-independent hopping along the $\alpha$ direction with $\alpha=x,y,z$, while $t_0=\Omega\int d{\mathbf{r}}w^*_{\mathbf{i}}({\mathbf{r}})|M_{\alpha'}(r)|w_{\mathbf{i}+\mathbf{1}_{\alpha'}}({\mathbf{r}})$ is the strength of spin-flipping hopping with $\alpha'=x,y$, where $w_{\mathbf{j}}({\mathbf{r}})\equiv w({\mathbf{r}}-{\mathbf{r}}_{\mathbf{j}})$ is the localized Wannier function of the lowest $s$ orbit.
And $\mathcal{R}_{z}=\exp(-i\sqrt{2}\pi\hat{F}_z)$ is induced by the Peierls substitution with Peierls phase $\sqrt{2}k_La=\sqrt{2}\pi$.

After the gauge transformation, $\hat{a}_{\mathbf{j},0}\rightarrow (-1)^{(m+n)}i\hat{a}_{\mathbf{j},0}$, we obtain the lattice Hamiltonian provided in the main text,
\begin{eqnarray}\label{Eq.realspacelat}
{H_0}=\!-\!\!&&\sum_{\mathbf{j},\alpha=x,y}[it_0\hat{\psi}^{\dag}_{\mathbf{j}}\hat{F}_{\alpha}\hat{\psi}_{\mathbf{j} +\mathbf{1}_\alpha} \!+t\hat{\psi}^{\dag}_{\mathbf{j}}(2\hat{F}^2_{z}-\hat{I})\hat{\psi}_{\mathbf{j}+\mathbf{1}_{\alpha}} \!+ {\rm H.c.}]
\nonumber \\
-&&t\sum_{\mathbf{j}}(\hat{\psi}^{\dag}_{\mathbf{j}}\mathcal{R}^{\sigma\sigma}_{z}\hat{\psi}_{\mathbf{j}+\mathbf{1}_z}+ {\rm H.c.})
\nonumber \\
+&&\sum_{\mathbf{j}}[(\delta+\delta_2)\hat{n}_{\mathbf{j},\uparrow}-(\delta-\delta_2)\hat{n}_{\mathbf{j},\downarrow}],
\end{eqnarray}
with $\hat{\psi}_{\mathbf{j}}=[\hat{a}_{\mathbf{j},\uparrow},\hat{a}_{\mathbf{j},0},\hat{a}_{\mathbf{j},\downarrow}]^T$.

\section{The effective Hamiltonian of WPs\label{appB}}
To reveal the topological features of twofold WPs more directly, we derive their effective two-band Hamiltonian in this section. We will demonstrate that the effective two-band Hamiltonian of single WPs is made up of linear spin-vector-momentum coupling, which is quadratic dispersion along the $x$ and $y$ directions for double WPs. Without loss of generality, we just calculate the results about the single WP with momentum $(0,0,k_{SW}=0.08\pi/a)$ and double WP with momentum $[0,0,k_{DW}=2\pi/(3a)]$ for the conventional single TDP phase, respectively.

The Hamiltonian of the single WP in the low-energy limit with neglecting the higher-order dispersion takes the form as (with ignoring a constant term)
\begin{align}\label{Eq.effecWP}
H_{SW}(\mathbf{q}) &= \sum_{\alpha=x,y} 2t_0aq_\alpha\hat{F}_{\alpha}-t_1\cos(k_{SW}a)aq_z\hat{F}_{z}
\nonumber \\&+ t_2\sin(k_{SW}a)aq_z\hat{F}^2_{z}
\nonumber \\
&+ t_1[\sin(\pi/3)-\sin(k_{SW}a)](\hat{F}_{z}-\hat{F}^2_{z}),
\end{align}
where $\mathbf{q}$ indicates the wave vector with respect to the location of single WP $(0,0,k_{SW}=0.08\pi/a)$, $t_1=2t\sin(\sqrt{2}\pi)$, and $t_2=2t[\cos(\sqrt{2}\pi)-1]$. Correspondingly, the Heisenberg equation reads
\begin{eqnarray}\label{Eq.HeisenbergWP}
i\frac{\partial\hat{b}_{\uparrow}}{\partial t}&=&[t_2\sin(k_{SW}a)-t_1\cos(k_{SW}a)]aq_z\hat{b}_{\uparrow}\nonumber \\
&&+\sqrt{2}t_0a({q_x-iq_y})\hat{b}_{0}, \nonumber \\
i\frac{\partial\hat{b}_{0}}{\partial t}&=&\sqrt{2}t_0a[({q_x+iq_y})\hat{b}_{\uparrow}+({q_x-iq_y})\hat{b}_{\downarrow}], \nonumber \\
i\frac{\partial\hat{b}_{\downarrow}}{\partial t}&=&\sqrt{2}t_0a({q_x+iq_y})\hat{b}_{0}\nonumber \\&&+[t_2\sin(k_{SW}a)+t_1\cos(k_{SW}a)]aq_z\hat{b}_{\downarrow}\nonumber \\&&-2t_1[\sin(\pi/3)-\sin(k_{SW}a)]\hat{b}_{\downarrow}.
 \end{eqnarray}
Obviously, the energies of spin-$\uparrow$ and spin-$0$ components equal zero when $\mathbf{q}=0$,
so these two spin components degenerate at this single WP. Then we can adiabatically eliminate the spin-down component to get the effective two-band Hamiltonian, and one may set $i\dot{\hat{b}}_{\downarrow}=0$. As a result, we have
\begin{eqnarray}\label{Eq.eliminateWP}
\hat{b}_{\downarrow}\approx\frac{\sqrt{2}t_0a({q_x+iq_y})}{2t_1[\sin(\pi/3)-\sin(k_{SW}a)]}\hat{b}_{0},
 \end{eqnarray}
where we ignore the term in proportion to $q_z$ in the denominator under the limit $\mathbf{q}\rightarrow0$. By applying the condition $t_0=t$, Eq.~(\ref{Eq.eliminateWP}) can be simplified to
\begin{eqnarray}\label{Eq.eliminateWPsim}
\hat{b}_{\downarrow}\approx-0.49({q_x+iq_y})a\hat{b}_{0}.
 \end{eqnarray}

After substituting Eq.~(\ref{Eq.eliminateWPsim}) to Eq.~(\ref{Eq.HeisenbergWP}), we obtain the effective Heisenberg equation about degenerating states,
\begin{eqnarray}\label{Eq.HeisenTwoWP}
i\frac{\partial\hat{b}_{\uparrow}}{\partial t}&=&[t_2\sin(k_{SW}a)-t_1\cos(k_{SW}a)]aq_z\hat{b}_{\uparrow}\nonumber \\&&+\sqrt{2}t_0a({q_x-iq_y})\hat{b}_{0}, \nonumber \\
i\frac{\partial\hat{b}_{0}}{\partial t}&=&\sqrt{2}t_0a[({q_x+iq_y})\hat{b}_{\uparrow}-0.49({q_x+iq_y})a\hat{b}_{0}]. \end{eqnarray}
After neglecting the unit term and higher-order dispersions, we can get the effective two-band Hamiltonian describing the low-energy excitation near this single WP,
\begin{eqnarray}\label{Eq.effectWP}
H(\mathbf{q})=&&\sqrt{2}t_0a\sum_{\alpha=x,y} q_{\alpha}\hat{\sigma}_{\alpha}\nonumber \\&&+\frac{[t_2\sin(k_{SW}a)-t_1\cos(k_{SW}a)]aq_z}{2}\hat{\sigma}_{z},
\end{eqnarray}
where $\hat{\sigma}_{\alpha}$ indicates the Pauli matrix.

Obviously, the Hamiltonian~(\ref{Eq.effectWP}) is dominated by the linear dispersions and describes a singly WP with topological charge $\mathcal{C}=1$ considering the coefficient of the second term $[t_2\sin(k_{SW}a)-t_1\cos(k_{SW}a)]/{2}\approx0.62>0$. We should note that the other singly WPs have similar results.

In contrast, the Hamiltonian of the double WP with momentum $[0,0,k_{DW}=2\pi/(3a)]$ in the low-energy limit is
\begin{eqnarray}\label{Eq.effecDWP}
H_{DW}(\mathbf{q})=&&2t_0a\sum_{\alpha=x,y}q_\alpha\hat{F}_{\alpha}-t_1\cos(k_{DW}a)aq_z\hat{F}_{z}
\nonumber \\&&+t_2\sin(k_{DW}a)aq_z\hat{F}^2_{z}+t_2(\hat{F}^2_{z}-\hat{I}),
\end{eqnarray}
where $\mathbf{q}$ indicates the wave vector with respect to this double WP. Then the corresponding effective Heisenberg equation reads
\begin{eqnarray}\label{Eq.HeisenbergDWP}
i\frac{\partial\hat{b}_{\uparrow}}{\partial t}&=&[t_2\sin(k_{DW}a)-t_1\cos(k_{DW}a)]aq_z\hat{b}_{\uparrow}\nonumber \\&&+\sqrt{2}t_0a({q_x-iq_y})\hat{b}_{0}, \nonumber \\
i\frac{\partial\hat{b}_{0}}{\partial t}&=&\sqrt{2}t_0a[({q_x+iq_y})\hat{b}_{\uparrow}+({q_x-iq_y})\hat{b}_{\downarrow}]-t_2\hat{b}_{0}, \nonumber \\
i\frac{\partial\hat{b}_{\downarrow}}{\partial t}&=&\sqrt{2}t_0a({q_x+iq_y})\hat{b}_{0}\nonumber \\&&+[t_2\sin(k_{DW}a)+t_1\cos(k_{DW}a)]aq_z\hat{b}_{\downarrow}.
 \end{eqnarray}
In contrast to single WPs, the spin-$\uparrow$ and -$\downarrow$ components degenerate at this double WP (the energies of these two spin components equal to zero when $\mathbf{q}=0$). We can get the effective two-band Hamiltonian through adiabatically eliminating the spin-$0$ component, i.e., $i\dot{\hat{b}}_{0}=0$, which yields
\begin{eqnarray}\label{Eq.eliminate}
\hat{b}_{0}=\frac{\sqrt{2}t_0a}{t_2}[(q_x+iq_y)\hat{b}_{\uparrow}+(q_x-iq_y)\hat{b}_{\downarrow}],
 \end{eqnarray}
and by using $t_0=t$, we have
\begin{eqnarray}\label{Eq.eliminateDWsim}
\hat{b}_{0}\approx-0.56a[(q_x+iq_y)\hat{b}_{\uparrow}+(q_x-iq_y)\hat{b}_{\downarrow}].
\end{eqnarray}

Substituting Eq.~(\ref{Eq.eliminateDWsim}) into Eq.~(\ref{Eq.HeisenbergDWP}), one can derive an effective Heisenberg equation for the degenerating space:
\begin{eqnarray}\label{Eq.HeisenTwoDWP}
i\frac{\partial\hat{b}_{\uparrow}}{\partial t}&=&[t_2\sin(k_{DW}a)-t_1\cos(k_{DW}a)]aq_z\hat{b}_{\uparrow}\nonumber \\
&&-0.79t_0a^2[(q^2_x+q^2_y)\hat{b}_{\uparrow}+(q_x-iq_y)^2\hat{b}_{\downarrow}], \nonumber \\
i\frac{\partial\hat{b}_{\downarrow}}{\partial t}&=&-0.79t_0[(q_x+iq_y)^2\hat{b}_{\uparrow}+(q^2_x+q^2_y)\hat{b}_{\downarrow}]\nonumber \\&&+[t_2\sin(k_{DW}a)+t_1\cos(k_{DW}a)]aq_z\hat{b}_{\downarrow}.
\end{eqnarray}
As a result, the effective two-band Hamiltonian for the double WP reads
\begin{eqnarray}\label{Eq.effectDWP}
H(\mathbf{q})=&&-0.79t_0[({q^2_x-q^2_y})\hat{\sigma}_{x}+2{q_xq_y}\hat{\sigma}_{y}]\nonumber \\&&-t_1\cos(k_{DW}a)aq_z\hat{\sigma}_{z}.
\end{eqnarray}
As can be seen, the obtained  Hamiltonian is dominated by the quadratic dispersions along $x$ and $y$ directions, in contrast to the  Hamiltonian~(\ref{Eq.effectWP}). We find that this double WP carries a double topological charge with $\mathcal{C}=-2$.

\section{Surface spectral function\label{appC}}
\begin{figure}[htp]
\includegraphics[width=1.0\columnwidth]{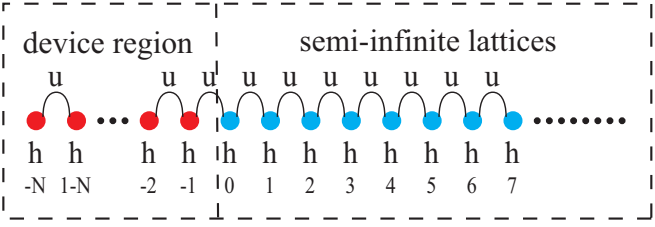}
\caption{\label{sup_Green_fun}The structure describing the effective one-dimensional chain.}
\end{figure}

In this section, we present how to get the surface spectral function by using the recursive Green¡¯s function method~\cite{Sancho_1985}. To gain more insight into the topologically protected Fermi arcs, we consider the surface states of the system by imposing a hard-wall confinement along the $y$ direction (open boundary condition) direction. After the Fourier transformation along $x$ and $z$ direction (periodic boundary conditions), the effective one-dimensional Hamiltonian can be obtained,
\begin{eqnarray}\label{Eq.oneD}
{H_y}=&&\sum_{j}h\hat{\varphi}^{\dag}_{j}\hat{\varphi}_{j}+(u\hat{\varphi}^{\dag}_{j}\hat{\varphi}_{j+1}+{\rm H.c.}),
\end{eqnarray}
where $\hat{\varphi}_{j}=[\hat{a}_{j,\uparrow},\hat{a}_{j,0},\hat{a}_{j,\downarrow}]^T$ is the annihilate operator in the $j$th site along the $y$ direction, $h$ is the effective local Hamiltonian, and $u$ represents the effective nearest-neighbor intersite connection (see Fig.~\ref{sup_Green_fun}). Explicitly, we have
\begin{eqnarray}\label{Eq.operators}
h=&&2t_0\sin(k_xa)\hat{F}_x+[2t\cos(k_xa)-2t\cos(k_za)]\hat{I}\nonumber
\\&&+[\delta_2-4t\cos(k_xa)-t_2\cos(k_za)]\hat{F}^2_z\nonumber
\\&&+[\delta-t_1\sin(k_za)]\hat{F}_z,
\nonumber \\
u=&&-it_0\hat{F}_y-2t\hat{F}^2_z+t\hat{I}. \nonumber
\end{eqnarray}

To get the surface spectral function about the $(0\overline{1}0)$ surface, we first divide the system to the device region and semi-infinite lattices, which are labeled by red dots and blue dots, as shown in Fig.~\ref{sup_Green_fun}. Explicitly, the device Green function satisfies
\begin{eqnarray}\label{Eq.device_green}
G_D(E)=[E^{+}-\hat{H}_D-\Sigma_R(E)]^{-1},
\end{eqnarray}
where the $E^{+}=E+i\eta$, $\hat{H}_D$ is the Hamiltonian describing the uncoupling device region and $\Sigma_R(E)$ is the self-energy matrix describing the effects of the semi-infinite lattices, which can be obtained through the surface Green function $G_{0,0}$,
\begin{eqnarray}\label{Eq.self_energy}
\Sigma_R(E)={\cal U}G_{0,0}{\cal U}^{\dag},
\end{eqnarray}
where the operator ${\cal U}$ describes the coupling between the device region and the semi-infinite lattices. Once the device Green function is obtained, the surface spectral function can be calculated,
\begin{eqnarray}\label{Eq.LDOS}
\rho(E)=-\frac{1}{\pi}\Im[\sum_{s}G_D(s,s;E)],
\end{eqnarray}
where $s$ indicates sites belonging to the surface of the device region (left end). Thus the essential step is to calculate the surface Green function in our system.

Furthermore, we will use the recursive Green function method to calculate the surface Green function $G_{0,0}$~\cite{Sancho_1985}. The Green function for the semi-infinite lattices satisfies the equation
\begin{eqnarray}\label{Eq.deltagreen}
[(E^+-H_R)G(E)]_{n,m}=\delta_{n,m},
\end{eqnarray}
where $H_R$ is defined over semi-infinite lattices and can be divided to local Hamiltonian $h$ and connection operators $u$. Then a sequence of results can be obtained,
\begin{eqnarray}\label{Eq.sequence}
(E^{+}-h)G_{0,0}=&&I+uG_{1,0}\nonumber \\
(E^{+}-h)G_{1,0}=&&uG_{2,0}+u^{\dag}G_{0,0}\nonumber \\
\vdots \nonumber \\
(E^{+}-h)G_{n,0}=&&uG_{n+1,0}+u^{\dag}G_{n-1,0},
\end{eqnarray}
which yields
\begin{eqnarray}\label{Eq.sequence2}
[E&&^{+}-h-u(E^{+}-h)^{-1}u^{\dag}]G_{0,0}=I+u(E^{+}-h)^{-1}uG_{2,0}, \nonumber \\ &&[E^{+}-h-u(E^{+}-h)^{-1}u^{\dag}-u^{\dag}(E^{+}-h)^{-1}u]G_{n,0}, \nonumber \\
&&~~=u(E^{+}-h)^{-1}uG_{n+2,0}+u^{\dag}(E^{+}-h)^{-1}u^{\dag}G_{n-2,0}. \nonumber \\
\end{eqnarray}

To process further, a new recursion equation about the event site reads
\begin{eqnarray}\label{Eq.sequence3}
(E^{+}-\varepsilon^s_1)G_{0,0}=&&I+\alpha_1G_{2,0}\nonumber \\
(E^{+}-\varepsilon_1)G_{2,0}=&&\alpha_1G_{4,0}+\beta_1G_{0,0}\nonumber \\
\vdots \nonumber \\
(E^{+}-\varepsilon_1)G_{2n,0}=&&\alpha_1G_{2(n+1),0}+\beta_1G_{2(n-1),0},
\end{eqnarray}
with
\begin{eqnarray}\label{Eq.parameter1}
\alpha_1=&&u(E^{+}-h)^{-1}u,\nonumber \\
\beta_1=&&u^{\dag}(E^{+}-h)^{-1}u^{\dag},\nonumber \\
\varepsilon^s_1=&&h+u(E^{+}-h)^{-1}u^{\dag},\nonumber \\
\varepsilon_1=&&h+u^{\dag}(E^{+}-h)^{-1}u. \nonumber
\end{eqnarray}
By repeating this process $k$ times, one can derive the following equations:
\begin{eqnarray}\label{Eq.sequence4}
(E^{+}-\varepsilon^s_k)G_{0,0}=&&I+\alpha_kG_{2^k,0}\nonumber \\
(E^{+}-\varepsilon_k)G_{2^k,0}=&&\alpha_kG_{2^k\cdot2,0}+\beta_kG_{0,0}\nonumber \\
\vdots \nonumber \\
(E^{+}-\varepsilon_k)G_{2^k\cdot n,0}=&&\alpha_kG_{2^k\cdot(n+1),0}+\beta_kG_{2^k\cdot(n-1),0},\nonumber \\
\end{eqnarray}
with
\begin{eqnarray}\label{Eq.parameter2}
\alpha_k=&&\alpha_{k-1}(E^{+}-\varepsilon_{k-1})^{-1}\alpha_{k-1},\nonumber \\
\beta_k=&&\beta_{k-1}(E^{+}-\varepsilon_{k-1})^{-1}\beta_{k-1},\nonumber \\
\varepsilon^s_k=&&\varepsilon_{k-1}+\alpha_{k-1}(E^{+}-\varepsilon_{k-1})^{-1}\beta_{k-1},\nonumber \\
\varepsilon_k=&&\varepsilon^s_{k-1}+\beta_{k-1}(E^{+}-\varepsilon_{k-1})^{-1}\alpha_{k-1}.\nonumber \end{eqnarray}
When $||\alpha_k||$ and $||\beta_k||$ are sufficiently small, the surface Green function can be obtained via
\begin{eqnarray}\label{Eq.Surface_green}
G_{0,0}\approx(E^+-\varepsilon^s_k)^{-1}.
\end{eqnarray}

\begin{figure*}[htp]
\includegraphics[width=1.9\columnwidth]{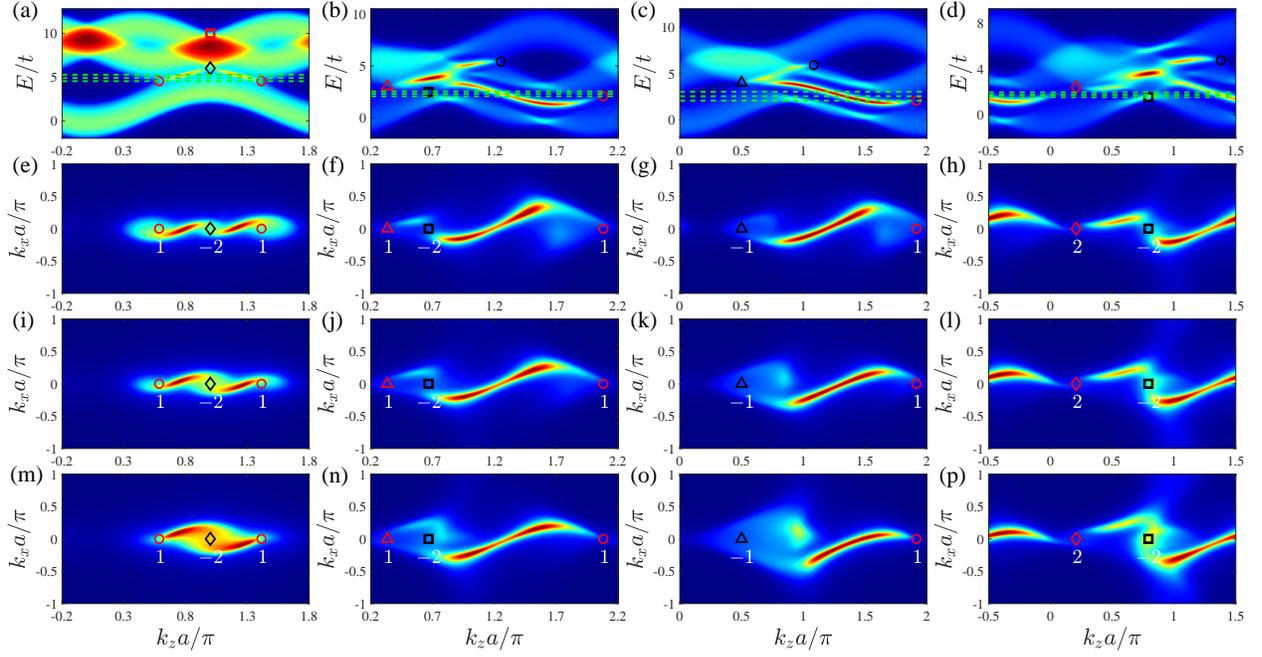}
\caption{\label{other_energy_fermi}(a)--(d) The surface spectral density of the $(0\overline{1}0)$ surface at the $k_z$-$E$ plane with $k_x=0$ corresponding to the parameters used in Figs.~\ref{TDP}(b)--\ref{TDP}(e) of the main text. (e)--(p) The surface spectral density at the surface Brillouin zone with the Fermi energy from the vicinity of TDP to WP shown as green dashed lines in (a)--(d). The red- (black-) circle (square) and triangle (diamond) denote the singly (doubly) quantized WPs and TDP with a positive (negative) topological charge, respectively. The white numbers represent topological charge of BDPs.}
\end{figure*}

In summary, the surface spectral function can be obtained through the following steps: (\uppercase\expandafter{\romannumeral1})~Using the recursion process Eqs.~(\ref{Eq.sequence3})--(\ref{Eq.parameter2}) to get $\varepsilon^s_k$. (\uppercase\expandafter{\romannumeral2})~Using Eq.~({\ref{Eq.Surface_green}}) to obtain the surface Green function. (\uppercase\expandafter{\romannumeral3})~Using Eqs.~({\ref{Eq.device_green}}) and ~({\ref{Eq.self_energy}}) to yield the device Green function. (\uppercase\expandafter{\romannumeral4})~Using Eq.~({\ref{Eq.LDOS}}) to calculate the surface spectral function of $(0\overline{1}0)$ surface. In addition, the surface spectral function of the $(010)$ can also be calculated by using the same methods with putting the device region on the right end.

\section{Fermi arcs of TDPs\label{appD}}
In this section we present the details of the surface spectral density for different Fermi energies. Figures~\ref{other_energy_fermi}(a)--\ref{other_energy_fermi}(d) shows the surface spectral density at the $k_z$-$E$ plane for four types of TDP phases using the recursive Green's function, respectively. Obviously, there are several surface states with energy between the middle and upper band. In contrast to the surface states below the middle band just depending on Chern number $\mathcal{C}_{-1}$, the surface states above the middle band are determined by the sum of the lower and middle bands' Chern number $(\mathcal{C}_{-1}+\mathcal{C}_{0})$. In particular, surface states always link to the BDPs with opposite topological charges, corresponding to the Chern number equal to $|\mathcal{C}_{-1}+\mathcal{C}_{0}|$. As to the doubly conventional TDP, there are two surface states above the middle band which both link the conventional TDP and the double WP with $\mathcal{C}=2$. By contrast, there is only one surface state connecting the TDP and WP between the middle and higher band for the singly conventional TDP and unconventional TDP phases.

Figures~\ref{other_energy_fermi}(b)--\ref{other_energy_fermi}(d) display the evolution of the Fermi arc for different TDPs. We begin with Fig.~\ref{other_energy_fermi}(b) corresponding to $k_0a=\pi/3$, where there is a single conventional TDP with $\mathcal{C}=1$, one single WP with Chern number $\{1,-1,0\}$, and a double WP with $\{-2,2,0\}$ for the middle and lowest band. When $k_0$ becomes larger, corresponding to the coupling strength $v_z$ ($u_z$) becoming large (small), the momentum difference between the TDP and the double WP will become smaller. Finally, these two BDPs will be converged when $k_0a=\pi/2$. Therefore a singly unconventional TDP appears, corresponding with its Chern number $\{-1,2,-1\}=\{1,0,-1\}+\{-2,2,0\}$ and giving rise to the surface states shown in Fig.~\ref{other_energy_fermi}(c). Analogously, the single conventional TDP will be converged into a single WP when $k_0a=\pi/3$ decreases to $k_0a=\arctan{t_1/t_2}$, which induces the double unconventional TDP ($\{2,-1,-1\}=\{1,0,-1\}+\{1,-1,0\}$) with surface states shown in Fig.~\ref{other_energy_fermi}(d). These results reveal that the number of BDPs is different for conventional and unconventional TDP phases, which corresponds to the distinctly linking structure of the Fermi arcs.

To complete the results of surface spectral density as shown in Figs.~\ref{energy_cur_fermi}(i)--\ref{energy_cur_fermi}(l) of the main text, we further plot the Fermi arcs with Fermi energies closer to twofold WPs in Figs.~\ref{other_energy_fermi}(e)--\ref{other_energy_fermi}(p). As can be seen, when the Fermi energy is far away from the TDP but close to the WPs, the bulk Fermi surface projections near the TDP (WP) are growing (decreasing). Finally, we find that the Fermi arcs will link to the twofold WPs directly with tuning the Fermi energy, as shown in Figs.~\ref{other_energy_fermi}(m)--\ref{other_energy_fermi}(p).
%\bibliography{STMCmetal_ref}


\end{document}